\documentclass[preprint,proceedings]{rmaa}

\usepackage{paralist}

\usepackage{psfrag,color}

\SetYear{2004}
\SetConfTitle{Compact Binaries in the Galaxy and Beyond}

\title{Jets from Accreting White Dwarfs} 

\author{J. L. Sokoloski\altaffilmark{1,2},
S. J. Kenyon\altaffilmark{1}, C. Brocksopp\altaffilmark{3},
C. R. Kaiser\altaffilmark{4}, E. M. Kellogg\altaffilmark{1}}
\altaffiltext{1}{Harvard-Smithsonian CfA, 60 Garden St., Cambridge, MA
02138, USA.}
\altaffiltext{2}{NSF Astronomy \& Astrophysics postdoctoral fellow.}
\altaffiltext{3}{Mullard Space Science Laboratory, UCL, Surrey, UK.} 
\altaffiltext{4}{U. Southampton, Hampshire SO17 1BJ, UK.} 

\shortauthor{Sokoloski et al.}

\shorttitle{Jets from White Dwarfs}

\suppressfulladdresses

\listofauthors{J. L. Sokoloski}
\listofauthors{S. J. Kenyon}
\listofauthors{C. Brocksopp}
\listofauthors{C. R. Kaiser}
\listofauthors{E. M. Kellogg}
\indexauthor{Sokoloski, J. L.}
\indexauthor{Kenyon, S. J.}
\indexauthor{Brocksopp, C.}
\indexauthor{Kaiser, C. R.}
\indexauthor{Kellogg, E. M.}

\abstract{Collimated outflows from accreting white dwarfs
have a vital role to play in the study of astrophysical jets.}

\addkeyword{Accretion, Accretion Disks}
\addkeyword{Binaries: Symbiotic}
\addkeyword{Stars: Winds, Outflows}
\addkeyword{White Dwarfs}

\begin{document}
\maketitle

Observationally, jets are associated with systems where material is
accreted though a disk.  Theoretically, accretion disks provide the
foundation for many jet models.  Perhaps the best-understood of all
accretion disks are those in cataclysmic variable stars (CVs).  Since
the disks in other accreting white-dwarf (WD) binaries are probably
similar to CV disks (at least to the extent that one does not expect
complications such as advection-dominated flows, for example), with WD
accretors we have the advantage of a relatively good understanding of
the region from which the outflows probably originate.

There are three main classes of accreting WDs: CVs, supersoft X-ray
binaries (SSXBs), and symbiotic stars (see Table 1).  In terms of
binary separation and orbital period, they form a hierarchy: CVs are
the most compact, with orbital periods of the order of hours; SSXBs
are slightly wider, with orbital periods from hours to days; and
symbiotics are the widest, with orbital periods on the order of years.
The WD luminosities do not form such a simple hierarchy.  In CVs the
luminosity of the accreting WD is low, but in SSXBs and most
symbiotics it is high enough that it would be difficult to produce by
accretion alone.  Thus, some of the accreted fuel in SSXBs and most
symbiotics probably undergoes thermonuclear fusion burning in a shell
on the WD surface.  To produce quasi-steady shell burning, either a
time-averaged accretion rate of roughly $10^{-7}$ to $10^{-8}$
M$_\odot$/yr or a recent 'thermal pulse' of runaway nuclear shell
burning may be required.  High accretion-rate CVs show modestly
collimated disk winds, but they have not been observed to produce
narrow, collimated jets.  Both SSXBs and symbiotics, however, have.
Based on this meeting, hot central stars of planetary nebulae (PNe)
would make an appropriate fourth column in Table 1, as some PNe also
produce collimated bi-polar outflows.

Symbiotic stars are numerous, unambiguous accreting systems, and close
enough for their collimated outflows to be spatially resolved.  We
therefore concentrate on symbiotics.  The WD in a symbiotic is fed
from the wind of the red giant.  This wind forms a nebula that is
partially ionized by radiation from the hot WD.  Shock heating of
nebular material by collimated ejecta could help make symbiotic-star
jets observable.  Collimated outflows from symbiotics are often
transient, and some are associated with poorly-understood classical
symbiotic outbursts.

\begin{table}[!t]\centering
  \setlength{\tabnotewidth}{\columnwidth}
  \tablecols{4}
  \setlength{\tabcolsep}{1.0\tabcolsep}
\footnotesize
  \caption{Accreting White Dwarfs}
  \begin{tabular}{rccc}
    \toprule
     & Cataclysmic & Supersoft & Symbiotic \\
     & Variables & Sources & Stars\\
    \midrule
 Size: & Small & Medium & Large\\
 Donor: & Dwarf & Evolved & Giant \\
 $L_{WD}$\tablenotemark{a}: & Few & $\sim 10^4$ & $\sim 10^3\;$\tablenotemark{b} \\
 $\dot{M}$: & Low & High & High \\
 $\dot{M}$ Mech: & Stable & Unstable & Wind \\ 
 &  RLO\tablenotemark{c} &  RLO &  \\ 
 Jets? & NO & YES & YES \\
    \bottomrule
    \tabnotetext{a}{Units of $L_\odot$.\hspace{0.4cm}$^b$Several
symbiotics have $L_{WD} \sim$ few~$L_\odot$. \hspace{0.5cm}$^c$RLO is short
for Roche-lobe overflow.}
  \end{tabular}
\end{table}

Symbiotic jets range in size from tens of milliarcsec to tens of
arcsec, corresponding to physical sizes of tens to thousands of AU.
Inferred flow speeds are hundreds to thousands of km/s (although
evidence for relativistic electrons has also been found; e.g., Crocker
et al.~2001).  Evidence for collimated outflows has been found in
practically every type of symbiotic star -- e.g., in systems with and
without nuclear shell burning on the WD, in those with strong and weak
WD magnetic fields, in 'D-type' and 'S-type' systems with
Mira/non-Mira mass donors, in symbiotic recurrent novae and systems
with classical symbiotic outbursts.  At least ten of the roughly 200
known symbiotics have shown either spectroscopic or imaging (radio or
optical) evidence for collimated outflows.  However, given their
transient nature and small angular extent on the sky, it is likely
that many symbiotic jet ejections have gone undetected, and that the
fraction of symbiotics that produce jets is much higher than the
currently known 5\%.  We discuss three examples below.

\begin{figure}[!t]
  \includegraphics[width=7.9cm]{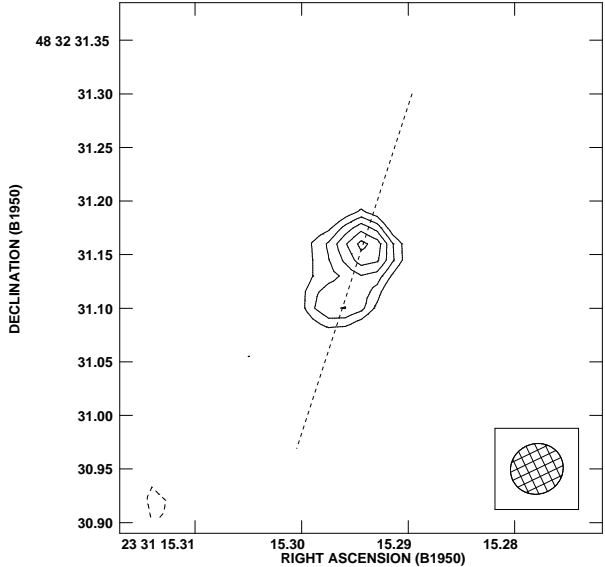}
  \caption{Transient extended radio structure in Z And.  The
  line is the projected perpendicular to the orbital plane.}
  \label{fig:zjet}
\end{figure}

To demonstrate the observational challenges faced in the study of WD
jets, we review the case of Z And.  The prototypical symbiotic star, Z
And has a strong WD magnetic field (Sokoloski \& Bildsten 1999), has a
high WD luminosity, and experiences classical symbiotic outbursts.
Brocksopp et al.~(2004) monitored Z And in the radio during the large
outburst beginning in 2000.  One year after the start of the optical
outburst, a small (60 mas) but significant extended radio structure
appeared (see Figure~\ref{fig:zjet}).  Within one month, the extended
structure had faded.  Because it was elongated perpendicular to the
orbital plane (see Schmid and Schild 1997), they interpreted the
transient structure as a jet-like ejection.  The outflow velocity was
at least 400 km/s, and the radio flux density indicated that the jet
emission was likely to have been thermal Bremsstrahlung.

Our second example, CH Cyg, highlights the link between WD accretors
and other types of jet-producing systems.  CH Cyg has a low enough WD
luminosity that it is unlikely to contain nuclear shell burning, and
it has large-amplitude CV-like optical flickering, presumably from an
accretion disk.  It moves between optical brightness states rather
than showing well-separated outbursts, and it tends to produce jets as
it changes between these states.  On one of the occasions of radio-jet
ejection (Karovska et al~1998), Sokoloski \& Kenyon (2003) found that
the most rapid flickering disappeared.  If the fastest variations come
from the region closest to the WD -- an idea supported by
eclipse-mapping studies in CVs and the theoretical expectation that
variability time scale is related to viscous and/or dynamical time
scales, both of which are shorter for smaller disk radii -- then it
appears that the inner disk was disrupted, or emission from the
inner-disk region suppressed when the jet was produced.  A similar
relationship between jet ejection and an change in the inner disk has
been found in some X-ray binaries.

Our final example, R Aqr, is the only WD currently known to have an
X-ray jet.  The X-ray emission from this nearby, Mira-containing,
long-period symbiotic was examined by Kellogg et al.~(2001).  At
energies above 1 keV, only the central point source was detected.  At
energies below 1 keV, they saw bi-polar structure with an X-ray
spectrum indicative of shock-heated plasma out of thermal equilibrium.
Although some X-ray features corresponded to features in the radio
map, the main X-ray bright spot in the northern jet was further from
the central object than the main radio bright spot.  Follow-up
$Chandra$ and VLA observations performed 3 years later, in 2003, show
significant changes in the jet structure in both the X-ray and radio
images (Kellogg et al.~2004).

Finding jets from WDs is a challenge.  Nonetheless, two out of three
classes of accreting WDs have shown evidence for jets (three out of
four if PNe are included).  The diversity of jet-producing WD systems
indicates that in some sense jets must be easy to produce.  One
element that could be common to all jet-producing WDs is an accretion
disk, perhaps with accretion rate or size above some threshold level.

\end{document}